\documentclass[article,12pt,reqno]{amsart} 
\usepackage{amsmath,amssymb}
\usepackage{color}

\theoremstyle{plain}
\newtheorem{thm}{Theorem}[section]

\theoremstyle{definition}

\usepackage{enumerate}

\newcommand{\bC}{{\mathbb C}}

\newcommand{\bR}{{\mathbb R}}

\newcommand{\bZ}{{\mathbb Z}}

\newcommand{\cD}{{\mathcal D}}

\renewcommand{\iint}{\int \kern -3pt\int}

\numberwithin{equation}{section}
\makeatletter
\@namedef{subjclassname@2020}{\textup{2020} Mathematics Subject Classification}
\makeatother

\newcommand{\nocontentsline}[3]{}
\let\origcontentsline\addcontentsline
\newcommand\stoptoc{\let\addcontentsline\nocontentsline}
\newcommand\resumetoc{\let\addcontentsline\origcontentsline}

\title[The Ozawa solution and surface theory]{The Ozawa solution to the Davey--Stewartson II equations and surface theory}

\author{Yi C. Huang}
\address{School of Mathematical Sciences, Nanjing Normal University, Nanjing 210023, People's Republic of China}
\email{Yi.Huang.Analysis@gmail.com}
\urladdr{https://orcid.org/0000-0002-1297-7674}

\author{Iskander A. Taimanov}
\address{Sobolev Institute of Mathematics, 630090 Novosibirsk, Russia}
\email{taimanov@math.nsc.ru}
\urladdr{http://www.math.nsc.ru/~taimanov/}

\date{\today}


\begin{document}

\begin{abstract}
We describe the Ozawa solution to the Davey--Ste\-wart\-son II equation from the point of view of surface theory by presenting a soliton deformation of surfaces which is ruled by the Ozawa solution. The Ozawa solution blows up at certain moment and we describe explicitly  the corresponding singularity of the deformed surface.
\end{abstract}

\maketitle

\hfill{To V.V. Kozlov on his 75th birthday}


\section{Introduction}

The Davey--Stewartson II (DSII) equation:
\begin{equation}
\label{dsii}
U_t = i(U_{zz}+U_{\bar{z}\bar{z}} + (V+\bar{V})U),  \ \ \
V_{\bar{z}} = 2(|U|^2)_z,
\end{equation}
was introduced in \cite{DS} for describing certain surface waves. It is represented in the form of Manakov's $L,A,B$-triple
\begin{equation}
\label{lab}
\cD_t = [\cD,A] + B\cD,
\end{equation}
where
\begin{equation}
\label{diracc}
\cD = \left(
\begin{array}{cc}
0 & \partial \\
-\bar{\partial} & 0
\end{array}
\right) + \left(
\begin{array}{cc}
U & 0 \\
0 & \bar{U}
\end{array}
\right)
\end{equation}
is the two-dimensional Dirac operator with the potential $U$ 
and $A$ and $B$ are differential operators.
Here $\partial = \frac{1}{2}\left(\frac{\partial}{\partial x} - i \frac{\partial}{\partial y}\right),
\bar{\partial} = \frac{1}{2}\left(\frac{\partial}{\partial x} + i \frac{\partial}{\partial y}\right)$ and
$x$ and $y$ are the Euclidean coordinates on the two-plane.

To every pair $\psi$ and $\varphi$ of solutions of the Dirac equations
$$
\cD \Psi=0, \ \ \cD^\vee \Phi = 0,
$$
where $\cD^\vee$ is obtained from $\cD$ by swapping $U$ and $\bar{U}$, there corresponds a surface 
\begin{equation}
\label{weier}
\begin{aligned}
S(\Phi,\Psi) &=  \int
i\left(\begin{array}{cc} \psi_1\overline{\varphi}_2 & -\overline{\psi}_2\overline{\varphi}_2 \\
\psi_1 \varphi_1 & -\overline{\psi}_2 \varphi_1 \end{array} \right) dz +
i\left(\begin{array}{cc} \psi_2\overline{\varphi}_1 & \overline{\psi}_1\overline{\varphi}_1 \\
- \psi_2 \varphi_2 & -\overline{\psi}_1\varphi_2 \end{array} \right) d\overline{z}\\
&
=\int d \left(\begin{array}{cc} ix^3 + x^4 & -x^1-ix^2 \\
x^1-ix^2 & -ix^3+x^4 \end{array}\right)
\end{aligned}
\end{equation}
defined up to translations \cite{Kon} and moreover every closed surface in
the four-space admits such a representation \cite{TDS}.
Therewith $z$ is a conformal parameter on the surface and 
$$
\int |U|^2 dx \wedge dy
$$
is the one fourth of the value of the Willmore functional on the surface.  The function $U$ is called the potential of a surface (with a fixed conformal parameter $z$) and it is defined up to gauge transformations \cite{T06}.

We refer for the general expositions of this so-called Weierstrass (spinor) representations of surfaces in $3$- and $4$-spaces, their applications and some open problems to \cite{T06,T23}.

In \cite{Kon} for surfaces ``induced'' by \eqref{weier} there were defined their soliton deformations. Since \eqref{dsii} admits the representation
\eqref{lab} their ``eigenfunctions'' $\Psi$ and $\Phi$ evolve
as
$$
\partial_t \Psi = A \Psi, \ \ \partial_t \Phi = A^\vee \Phi,
$$
where $A$ enters \eqref{lab} and $A^\vee$ is also a certain differential operator. 
That defines a deformation of the surface \eqref{weier}  which may be called a soliton deformation. 

For $U=\bar{U}$ and $\Phi=\Psi$ formulas \eqref{weier} describe surface in the three-space, i.e. $x^4=0$, and originally 
that was done in \cite{K1} for this situation. In \cite{T97} it was proved that every closed surface in $\bR^3$ admits such a representation
and the corresponding soliton deformation given by the modified Novikov--Veselov equation, preserves the Willmore functional.

In \cite{T21} soliton deformations were used for constructing singular solutions to the Davey--Stewartson II equation.
There was constructed a soliton deformation of minimal embeddings $\Gamma(t)$ of $\bC$ into $\bR^4$ and by the  transformation
$S \to S^{-1}$ these surfaces were mapped into branched Willmore spheres $\widetilde{\Gamma}(t)$. To the mapping $S \to S^{-1}$ there corresponds 
the Moutard transformation between solutions of the Davey--Stewartson equations \cite{MT}. In \cite{T21} it maps the stationary solution 
$U_t=0$ to nontrivial solution $\widetilde{U}$ such that

1)  $U$ is nonsingular if the initial minimal surface does not pass through the origin;

2) if  $\Gamma(t)$ passes through the origin at point $z=z_0$ for $t=t_0$, then this point is mapped to the infinity by the transformation $S \to S^{-1}$
and $\widetilde{U}$ has a singularity of the indeterminacy type:
$$
\widetilde{U} \sim a \cdot e^{b i \phi} \ \ \mbox{as $r \to 0$},
$$
where $a,b$ are real-valued nonzero constants and $z-z_0 = r e^{i\phi}$; 

3) $\int |\widetilde{U}|^2 \, dx \wedge dy$ is an integer multiple of $\pi$ (since, $\widetilde{\Gamma}(t)$ are Willmore spheres)
and it has the same value for nonsingular times and it changes by of type $k \pi, k \in \bZ$, at singular times. In particular, 
$\|\widetilde{U}\|^2 <\infty$, i.e. $\widetilde{U} \in L_2(\bR^2)$.

In \cite{T21} there were constructed solutions with just one such a singularity.

Before such type of solutions were constructed in \cite{T15} for the modified Novikov--Veselov equation by using the same trick applied to 
the Weierstrass representation of surfaces in $\bR^3$ \cite{K1,T97}.

Recently in \cite{Huang1} there were constructed solutions to the Davey--Ste\-wart\-son II equations 
with two indeterminacies at the same moment of time. 
In this case the surfaces $\Gamma(t)$ are not embedded. 

Until recently there are known two scenarios of creating singularities of solutions to the Davey--Stewartson II equation with initial Cauchy data from $L_2$. These are the one described above and leading to indeterminacies and the Ozawa solution discovered in 1992 \cite{Oza92}.  
It is
\begin{equation}
\label{ozawa}
\widetilde{U}(z,\bar{z},t) = \frac{1}{t}e^{-i\frac{z^2+\bar{z}^2}{8}} \frac{1}{1+ \frac{|z|^2}{t^2}}
\end{equation}
which is 

1) nonsingular for $t \neq 0$;

2) $|\widetilde{U}| \to 0$ for $z \neq 0$, $t \to 0$;

3) $\int |\widetilde{U}|^2 \, dx \wedge dy = \pi$ for $t \neq 0$ and
$$
|\widetilde{U}|^2 \to \pi \delta_0\ \ \mbox{as $t \to 0$},
$$
where $\delta_0$ is the delta distribution at the origin of $\bC$. 

It is a natural problem:

{\sl to present a soliton deformation of surfaces which is ruled by the Ozawa solution and demonstrate what is the geometrical counterpart of the
Ozawa singularity.}

In this article we solve it (see Theorem \ref{theo}).

\section{Surface theory for pseudo-conformal transformation}

Instead of the Moutard transformation mechanism,
we study the singularity formation for the Davey-Stewartson II equation via the pseudo-conformal transformation \cite{GV} 
that was used by Ozawa \cite{Oza92}.
We however take the geometric viewpoint.

Consider the following system of Dirac equations
$$
\cD \Psi = \left[\left(
\begin{array}{cc}
0 & \partial  \\
-\overline{\partial} & 0
\end{array}
\right) + \left(
\begin{array}{cc}
U & 0\\
0 & \overline U
\end{array}
\right)\right]\left(\begin{array}{c} \psi_1 \\ \psi_2 \end{array}\right)=0,
$$
$$
\cD^\vee \Phi = \left[\left(
\begin{array}{cc}
0 & \partial  \\
-\overline{\partial} & 0
\end{array}
\right) + \left(
\begin{array}{cc}
\overline U & 0\\
0 &  U
\end{array}
\right)\right]\left(\begin{array}{c} \varphi_1 \\ \varphi_2 \end{array}\right)=0,
$$
and the so-called pseudo-conformal transformation on the potential
$$
U(z,\overline{z},t)\mapsto\widetilde{U}(z,\overline{z},t)=\frac1te^{-i\frac{z^2+\overline{z}^2}{8t}}U\left(\frac{z}{t},\frac{\overline z}{t},\frac{-1}{t}\right),
$$
with the following reduction when $U$ is independent of the temporal variable,
$$
U(z,\overline{z})\mapsto\widetilde{U}(z,\overline{z},t)=\frac1te^{-i\frac{z^2+\overline{z}^2}{8t}}U\left(\frac{z}{t},\frac{\overline z}{t}\right).
$$

Here is our first observation:
$$
\widetilde{\psi}_1(z,\overline{z},t)=e^{i\frac{z^2}{8t}}\psi_1\left(\frac{z}{t},\frac{\overline z}{t}\right),
\quad\widetilde{\psi}_2(z,\overline{z},t)=e^{-i\frac{\overline{z}^2}{8t}}\psi_2\left(\frac{z}{t},\frac{\overline z}{t}\right)
$$
and
$$
\widetilde{\varphi}_1(z,\overline{z},t)=e^{-i\frac{z^2}{8t}}\varphi_1\left(\frac{z}{t},\frac{\overline z}{t}\right),
\quad \widetilde{\varphi}_2(z,\overline{z},t)=e^{i\frac{\overline{z}^2}{8t}}\varphi_2\left(\frac{z}{t},\frac{\overline z}{t}\right)
$$
solve the Dirac equations with $\widetilde U$ if $\Psi = \left(\begin{array}{c} \psi_1 \\ \psi_2 \end{array}\right)$
and $\Phi = \left(\begin{array}{c} \varphi_1 \\ \varphi_2 \end{array}\right)$ solve the Dirac equations with $U$.

Now, consider the $t$-independent real-valued radial potential $$U=\frac{1}{1+z\overline{z}}.$$
Its transformation $\widetilde U$ is the Ozawa blowup solution \cite{Oza92} with singularity time $t=0$ for the Davey-Stewartson II equation written in complex variables.
Indeed, since $\|\widetilde U(\cdot,t)\|_2^2$ is conserved ($= \|\widetilde U(\cdot,1)\|_2^2=\pi$), and $|\widetilde U(z,\overline z,t)|$ for $z\neq0$ tends to 0 as $t\rightarrow0$, we see that
$$
|\widetilde U(\cdot,t)|^2\longrightarrow \pi\delta_0\quad\text{as}\quad t\rightarrow0.
$$

Our second observation is that we can solve the Dirac equation for this $U$ with
$$\psi_1=\frac{\overline{z}}{1+z\overline{z}}=\varphi_1,\quad \psi_2=\frac{1}{1+z\overline{z}}=\varphi_2.$$
Applying above transformations to this stationary $U$ we get a complex-valued non-radial potential $\widetilde U$ (a solution of the Davey-Stewartson II equation),
with the following spinors
$$
\widetilde{\Psi}= \left(
\begin{array}{cc}
\frac{t\overline z}{t^2+|z|^2}e^{i\frac{z^2}{8t}} & -\frac{t^2}{t^2+|z|^2}e^{i\frac{z^2}{8t}}\\
\frac{t^2}{t^2+|z|^2}e^{-i\frac{\overline{z}^2}{8t}} & \frac{t z}{t^2+|z|^2}e^{-i\frac{\overline{z}^2}{8t}}
\end{array}
\right),
$$
$$
\widetilde{\Phi}= \left(
\begin{array}{cc}
\frac{t\overline z}{t^2+|z|^2}e^{-i\frac{z^2}{8t}} & -\frac{t^2}{t^2+|z|^2}e^{-i\frac{z^2}{8t}}\\
\frac{t^2}{t^2+|z|^2}e^{i\frac{\overline{z}^2}{8t}} & \frac{t z}{t^2+|z|^2}e^{i\frac{\overline{z}^2}{8t}}
\end{array}
\right).$$

As before, we compute the Davey-Stewartson II deformation of surface
$$
\begin{aligned}
S(\widetilde\Phi,\widetilde\Psi) &=  \int
i\left(\begin{array}{cc} \widetilde\psi_1\overline{\widetilde\varphi}_2 & -\overline{\widetilde\psi}_2\overline{\widetilde\varphi}_2 \\
\widetilde\psi_1 \widetilde\varphi_1 & -\overline{\widetilde\psi}_2\widetilde\varphi_1 \end{array} \right) dz +
i\left(\begin{array}{cc}\widetilde \psi_2\overline{\widetilde\varphi}_1 & \overline{\widetilde\psi}_1\overline{\widetilde\varphi}_1 \\
- \widetilde\psi_2 \widetilde\varphi_2 & -\overline{\widetilde\psi}_1\widetilde\varphi_2 \end{array} \right) d\overline{z}\\
&=  \int
i\left(\begin{array}{cc} \frac{t^3\overline z}{(t^2+|z|^2)^2} & -\frac{t^4}{(t^2+|z|^2)^2} \\
\frac{t^2\overline z^2}{(t^2+|z|^2)^2} & -\frac{t^3\overline z}{(t^2+|z|^2)^2} \end{array} \right) dz +
i\left(\begin{array}{cc}\frac{t^3 z}{(t^2+|z|^2)^2} & \frac{t^2 z^2}{(t^2+|z|^2)^2} \\
-\frac{t^4}{(t^2+|z|^2)^2} & -\frac{t^3 z}{(t^2+|z|^2)^2} \end{array} \right) d\overline{z}\\
 &=\int d \left(\begin{array}{cc} \frac{-it^3 }{t^2+|z|^2} & \frac{-it^2 z}{t^2+|z|^2}  \\
\frac{-it^2\overline z}{t^2+|z|^2} & \frac{it^3 }{t^2+|z|^2}  \end{array} \right)=\int d \left(\begin{array}{cc} ix^3 + x^4 & -x^1-ix^2 \\
x^1-ix^2 & -ix^3+x^4 \end{array}\right).
\end{aligned}
$$

Thus we derive
\begin{thm}
\label{theo}
The Ozawa solution \eqref{ozawa} determines the potentials of surfaces in $\bR^3 \subset \bR^4$ 
from the $t$-parameter family 
given by the formulas
$$
w_1=x^1+ix^2=\frac{it^2 z}{t^2+|z|^2}=it\frac{z/t}{1+|z/t|^2},
$$
$$
w_2=x^3+ix^4=\frac{-t^3 }{t^2+|z|^2}=\frac{it}{z}w_1
$$
which lie in the hyperplane $x^4 = 0$.

For $z\neq0$, all the three coordinates shrink to 0 as $t\rightarrow0$.
For $z=0$, $x^1=x^2=0$, and the coordinate $x^3$ shrinks to 0 as $t\rightarrow0$.
\end{thm}

\section{Final remarks}

1) In \cite{T24} it was noticed that the singularities of the indeterminacy type for the solutions of the modified Novikov--Veselov equation \cite{T15}
and of the Davey--Stewartson equations \cite{T21} appear at the same time as the zero level of the discrete spectrum of $\cD$ changes
because certain eigenfunctions blow up. 

The Ozawa type singularity does not admit such an interpretation and in that respect also differs from indeterminacies. 
The ``good'' spectral theory for two-dimensional Dirac operators which, in particular, permits to develop the scattering theory
is constructed for potentials $U \in L_1 \cap L_\infty$ \cite{Sung}. 
However the Ozawa potentials are not integrable, i.e. do not lie in $L_1$.

2) It is interesting to notice that for all known examples of blowing up solutions of the Davey--Stewartson II equation the values
of the squared $L_2$-norms of potentials are quantized:
$$
\frac{1}{\pi} \int |U|^2\, dx \wedge dy \in \bZ.
$$
For  the cases of the indeterminacy that is because the deformed surfaces $\widetilde{\Gamma}(t)$ are branched Willmore spheres.
For the Ozawa case the reason is not known. Moreover Ozawa speculated in \cite{Oza92} that his solutions can be transformed in such 
a way that the above quantity could be as small as possible. However as it is mentioned in \cite{Huang2} this is not the case:
the parametric deformation of $L_2$-norms of potentials in \cite{Oza92} changed the magnitude of the nonlinearity in the Davey--Stewartson II equation.

\resumetoc

\vskip3mm

{\sc Acknowledgements.}

Research of the first author (Y.C.H.) is partially supported by the National NSF grant of China (no. 11801274), the
JSPS Invitational Fellowship for Research in Japan (no. S24040), and the Open Projects from Yunnan
Normal University (no. YNNUMA2403) and Soochow University (no. SDGC2418). 

Research of the second author (I.A.T.) was supported by the Mathematical Center in Academgorodok.

The work was done during 
the visit of the first author to the Sobolev Institute of Mathematics (January--February 2025).

\end{document}